\pdfoutput=1

\documentclass[3p,times]{elsarticle}

\usepackage{ecrc}

\volume{00}

\firstpage{1}

\journalname{Nuclear Physics A}

\runauth{G. O. V. de Barros, Bo Fenton-Olsen, Peter Jacobs, Mateusz P\l osko\'n}

\jid{nupha}

\usepackage{amssymb}
\usepackage{amsthm}

\usepackage[figuresright]{rotating}

\begin{document}

\begin{frontmatter}



\dochead{}

\title{Data-driven analysis methods for the measurement of reconstructed jets in heavy ion collisions at RHIC and LHC}


\author[USP]{G. O. V. de Barros}
\ead{gbarros@dfn.if.usp.br}
\author[LBL]{Bo Fenton-Olsen}
\author[LBL,CERN]{Peter Jacobs}
\author[LBL]{Mateusz P\l osko\'n}

\address[USP]{Universidade de S\~ao Paulo, Instituto de F\'isica\\
  Rua do Mat\~ao Travessa R, 187, 05508-090, S\~ao Paulo, Brazil}
\address[LBL]{Lawrence Berkeley National Laboratory\\
  1 Cyclotron Road, Berkeley, CA 94720, USA}
\address[CERN]{CERN\\
  CH-1211 Geneva 23, Switzerland}

\begin{abstract}

  We present data-driven methods for the full reconstruction of jets
  in heavy ion collisions, for inclusive and coincidence jet
  measurements at both RHIC and LHC. The complex structure of heavy
  ion events generates a large background of combinatorial jets,
  and smears the measured energy of the true hard jet
  signal. Techniques to correct for these background effects can induce
  biases in the reported jet distributions, which must be
  well controlled for accurate measurement of jet quenching. Using
  model studies, we evaluate the proposed
  methods for measuring jet distributions accurately while minimizing the
  fragmentation bias of the measured population.

\end{abstract}

\end{frontmatter}


\section{Introduction}
\label{sec:intro}

The interaction of QCD jets with the Quark-Gluon Plasma (QGP)
generated in high energy nuclear collisions (``jet quenching'')
provides unique and penetrating probes of the QGP. Jet quenching has
to date been measured primarily via the suppression of high $p_{\rm T}$
hadrons and their correlations \cite{Majumder:2010qh}. Full jet
reconstruction \cite{Aad:2010bu,Chatrchyan:2011sx} may enable the
study of jet quenching at the partonic level, without the added
complexity of hadronization. However, jet reconstruction in heavy ion
events is a challenging task, due to the large population of
combinatorial background jets generated by the jet reconstruction
algorithm from random recombination of particles not correlated via
single hard scattering, and to the distortion of the measured energy
of true jets by background fluctuations. Experimental control of these
effects may be achieved by pruning the soft component of events prior
to jet finding, and requiring jet candidates to contain a hard
fragmentation component. However, for quenched jets, these analysis
techniques may also bias the jet population and the reported jet
energy. For accurate quenching measurements it is necessary to assess
such biases and to minimize their effects. In these proceedings we present data-driven methods for
minimally-biased inclusive and coincidence jet measurements in heavy
ion collisions, at both RHIC and LHC. We base our approach on previous
developments
\cite{bib:pmj_hp10,bib:gbarros_PANIC11,Abelev:2012ej}. Background
fluctuations are ``unfolded'' using an iterative technique
incorporating Bayes's Theorem \cite{bib:D'Agostini}. We employ a model
event generator to evaluate these methods, and give prescriptions to
minimize fragmentation bias in real data analysis. This approach is
applicable to jet measurements by the STAR experiment at RHIC and
ALICE at the LHC, which are able to reconstruct a large fraction of
all jet fragments via precise charged particle tracking and
electro-magnetic calorimetry.

\section{Toy Model and Jet Reconstruction}
\label{sec:toymodel}
  
Our aim is to clarify the generic effects underlying the largest
systematic uncertainties in heavy ion jet measurements. For such a
parametric study, it is not necessary to model heavy ion events and
experimental response in detail; indeed, the complexity of such an
approach may obscure the important generic effects. We therefore
employ a simplified Toy Model event generator, that nevertheless
captures the main features of heavy ion jet reconstruction seen in
data \cite{bib:pmj_hp10,bib:gbarros_PANIC11,Abelev:2012ej}. The Toy Model randomly generates massless particles with uniform
distribution within $|\eta|<1$ and $0<\varphi<2\pi$. The transverse
momentum distribution of the soft background is a Boltzmann
distribution, with $\left<p_{\rm T}\right>=500$ MeV at RHIC and
$\left<p_{\rm T}\right>=700$ MeV at the LHC. The choice of jet distribution depends on
the observable: for inclusive jets we utilize the inclusive spectrum
measured in p+p collisions at the same $\sqrt{s}$, scaled by $T_{AA}$,
while for the coincidence measurement we utilize the semi-inclusive
distribution calculated by PYTHIA for p+p collisions, without
additional scaling. Jets are fragmented using the PYTHIA
fragmentation routines, or are not fragmented at all; i.e. a single
particle carries the total jet momentum (``SP'' fragmentation). The total
multiplicity per event corresponds to the average multiplicity of all
particles (charged plus neutral) in 0-5\% central Au+Au collisions at
RHIC or 0-5\% central Pb+Pb collisions at the LHC.

Millions of Toy Model event are generated, and are analyzed with the same
algorithms used for real data. Jets are reconstructed using the
Anti-$k_{T}$ algorithm \cite{bib:anti-kt} with resolution parameter
$R=0.4$ and energy recombination scheme, and with median background
density estimate $\rho$ as described in \cite{bib:bkg_sub}. We utilize
the \textsc{FASTJET} implementation with default settings
\cite{bib:fastjet}. Jet reconstruction incorporates all particles with
$p_{T} > 0.2$ GeV/c and $|\eta| < 1.0$, with full azimuthal coverage.
Accepted jets have $|\eta|<0.6$ and area $A_{\rm jet} > 0.4$
\cite{bib:pmj_hp10}. The measured jet energy $p_{T}^{\rm rec}$ is
corrected on an event-wise basis via \cite{bib:bkg_sub}:

\begin{equation}
  p_{T}^{\left<\rm corr\right>} = p_{T}^{\rm rec} - \rho\cdot A_{\rm jet},
  \label{eq:pTcorr}
\end{equation}

\noindent
where $\rho$ is the event-wise estimate of the background density
(energy per unit area) and
$\left<\rm corr\right>$ indicates event-wise correction for
background. Local variations of background density relative to $\rho$
will generate large distortions in the distribution of
$p_{T}^{\left<\rm corr\right>}$, which must be corrected via unfolding. We
use iterative unfolding incorporating Bayes's Theorem
\cite{bib:D'Agostini}, with the response matrix corresponding to
$\delta{p_{\rm T}}$, the distribution of fluctuations in jet response
measured using data \cite{bib:pmj_hp10,Abelev:2012ej}. In this
case, the ``data'' are Toy Model events.

\begin{figure}[!h]
  \centering
  \includegraphics[width=0.41\textwidth]{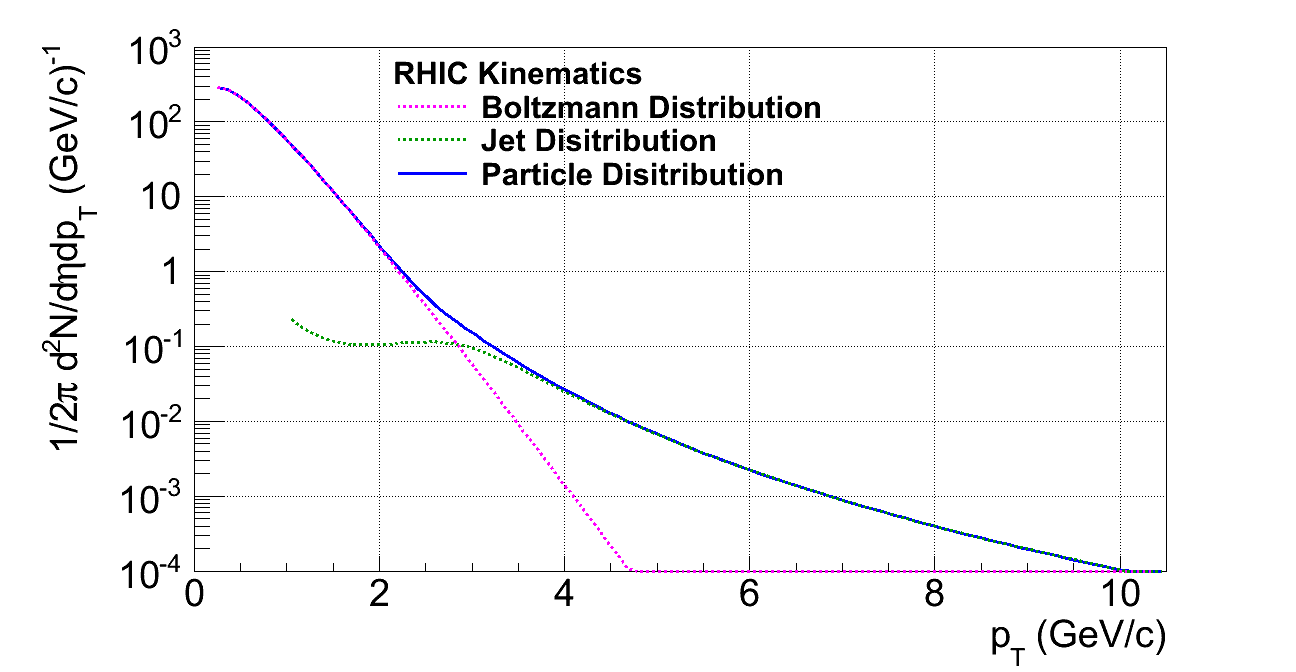}
  \includegraphics[width=0.41\textwidth]{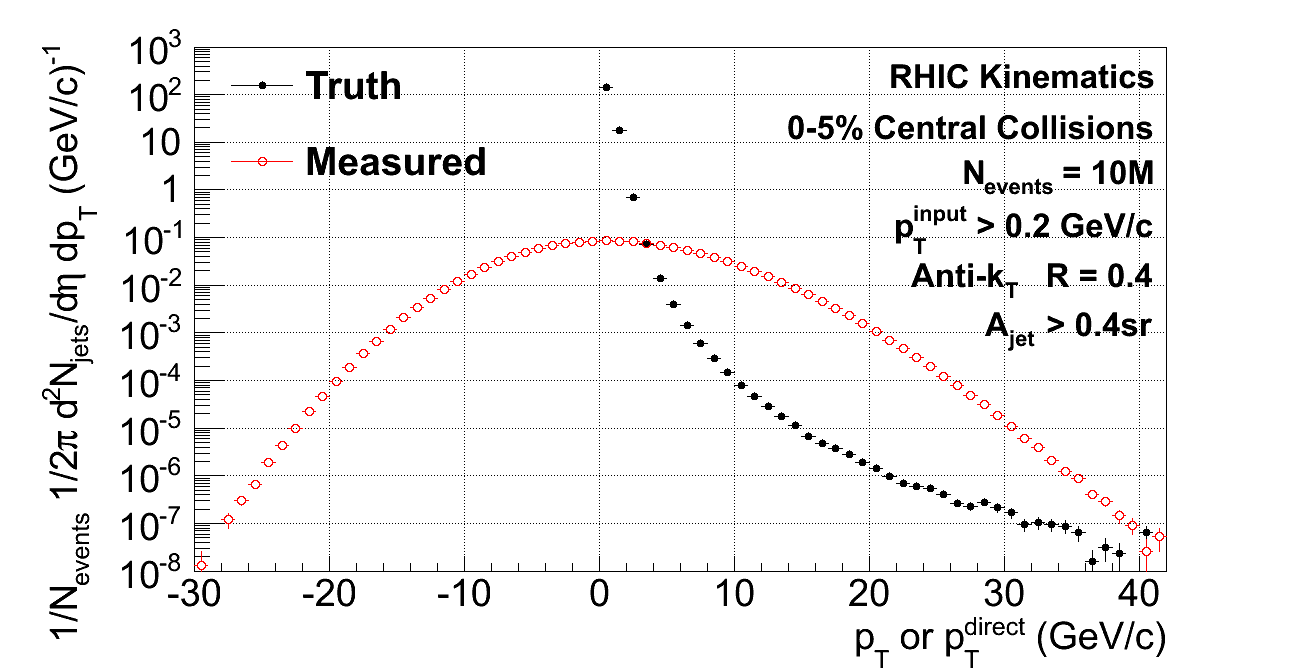}
  \caption{Toy Model distributions for central Au+Au collisions at
    RHIC. Left: particle distribution with SP fragmentation. Right:
    particle (``Truth'') and reconstructed jet spectra.}
  \label{fig:toy_dNdpT_pTcorr}
\end{figure}

The results presented here are for SP fragmentation, in order to
isolate the biases in the jet measurement due to suppression of the
combinatorial background component and unfolding of background
fluctuations. The validity of the SP approach is based on the observed
insensitivity of the response of the Anti-$k_{T}$ algorithm in heavy
ion events to the detailed pattern of fragmentation of particles
measured within the jet cone
\cite{bib:pmj_hp10,bib:gbarros_PANIC11}. Figure
\ref{fig:toy_dNdpT_pTcorr} shows the generated particle distribution
for SP fragmentation, together with the reconstructed
($p_{T}^{\left<\rm corr\right>}$) jet spectrum, for central Au+Au
collisions at RHIC.
 
\section{Inclusive Jet Spectrum Measurement}
\label{sec:inclusive}

\begin{figure}
  \centering
  \includegraphics[width=0.41\textwidth]{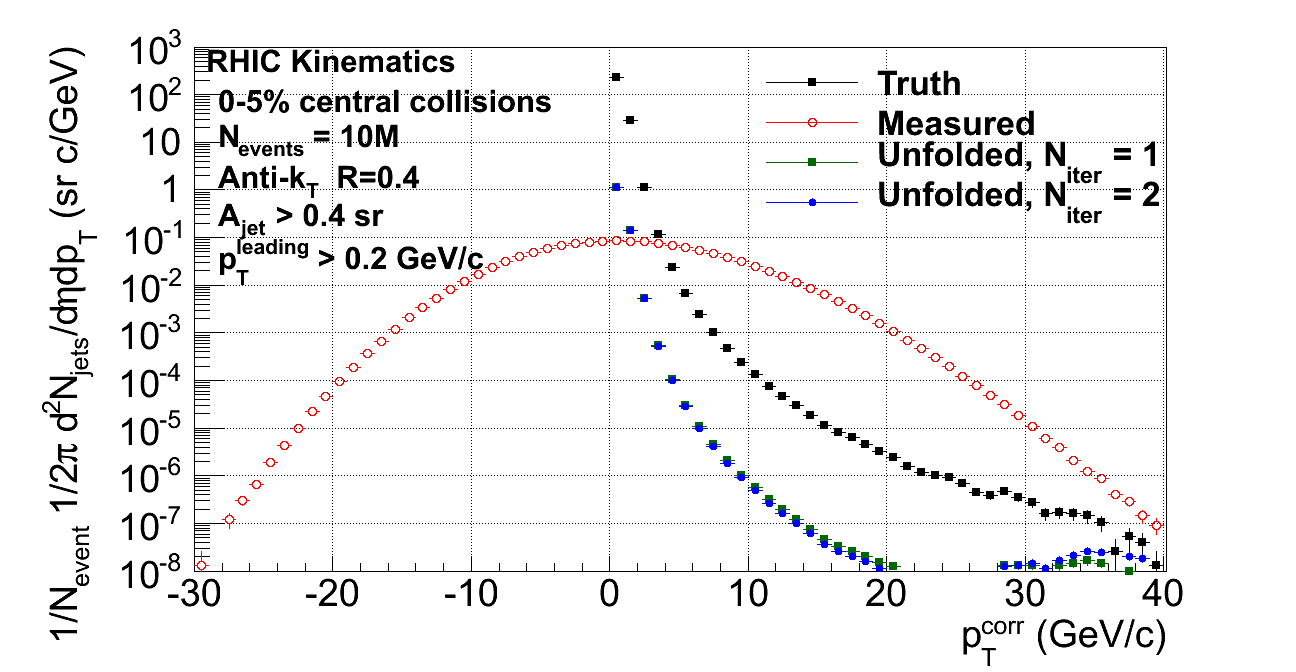}
  \includegraphics[width=0.41\textwidth]{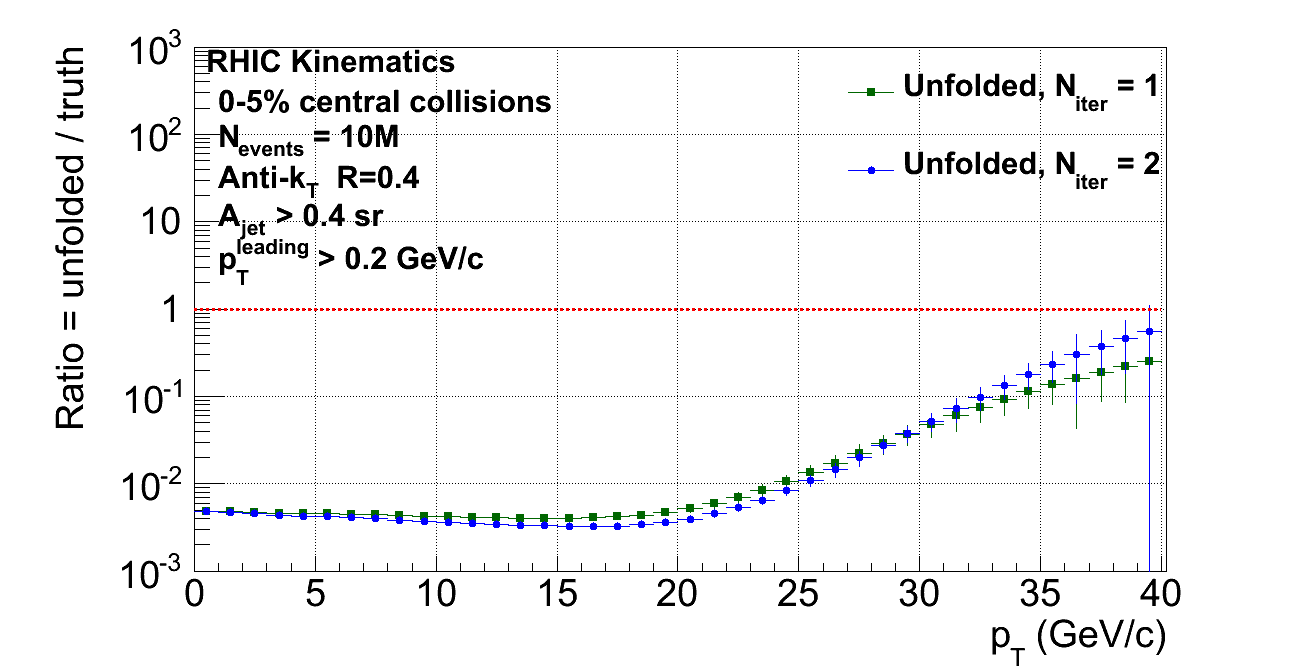}
  \includegraphics[width=0.41\textwidth]{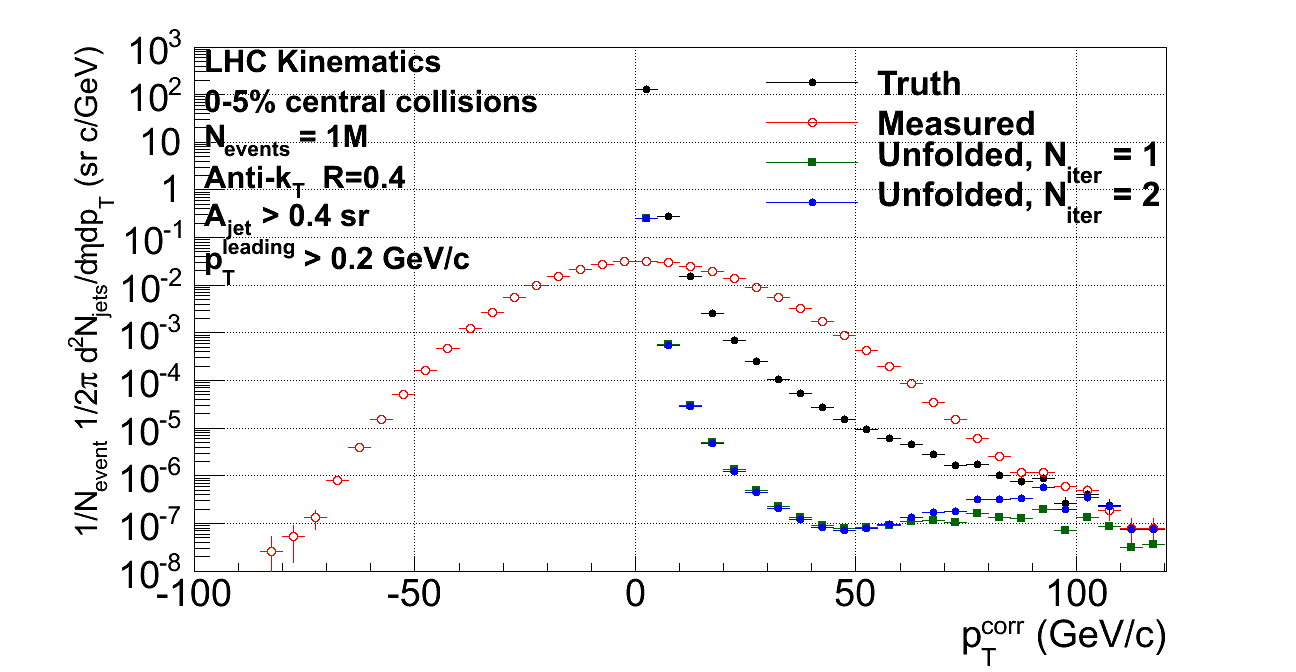}
  \includegraphics[width=0.41\textwidth]{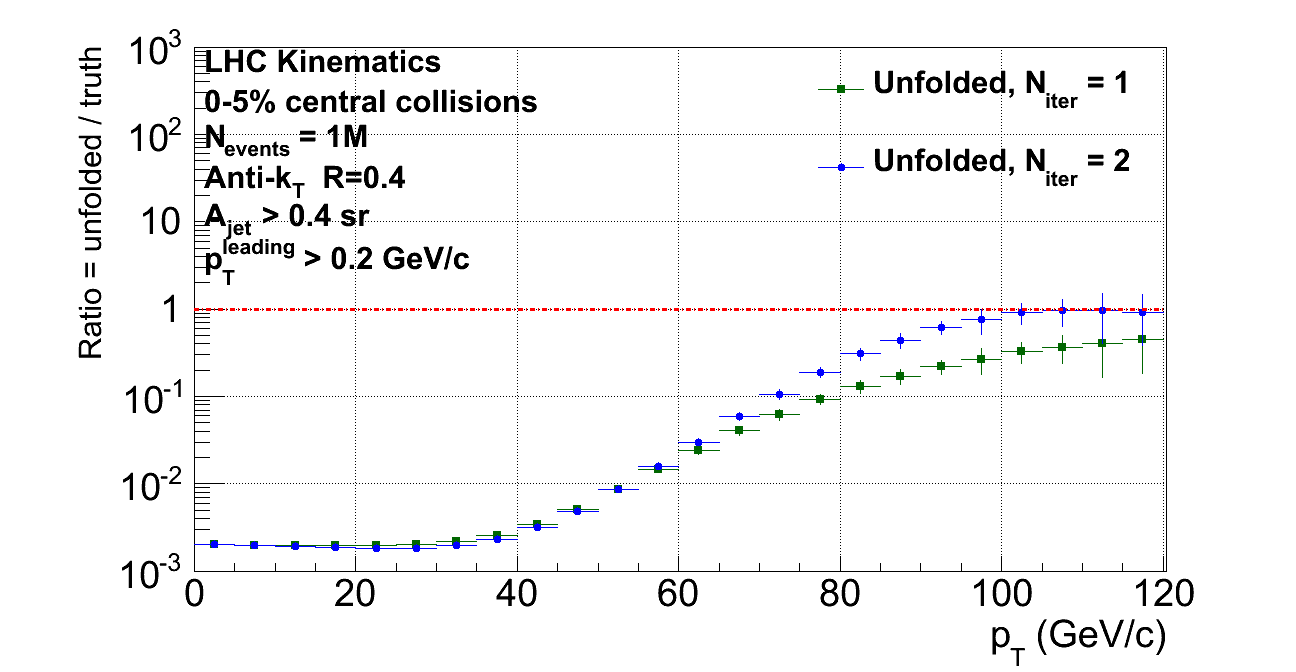}
  \caption{Measured, truth and unfolded distributions for fully
    inclusive jets. Upper: RHIC; lower: LHC. Right plots show ratio Unfolded/Truth.}
  \label{fig:unf_inc_nocut}
\end{figure}
  
\begin{figure}[!h]
  \centering
  \includegraphics[width=0.41\textwidth]{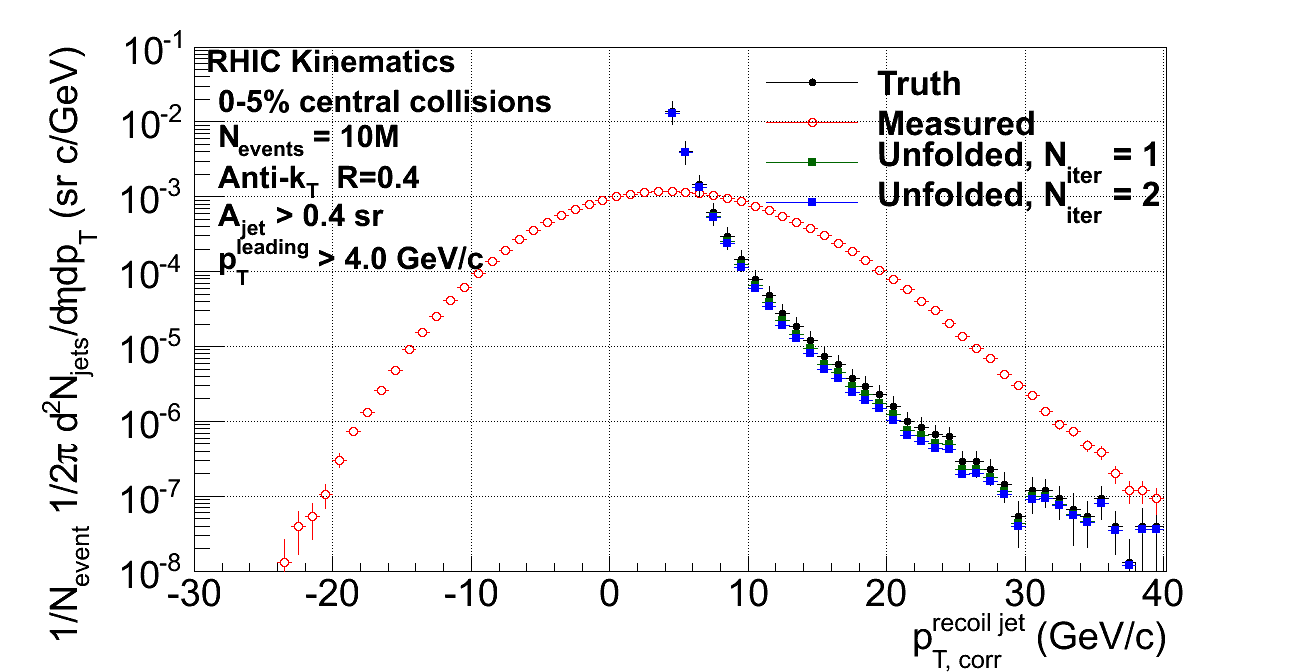}
  \includegraphics[width=0.41\textwidth]{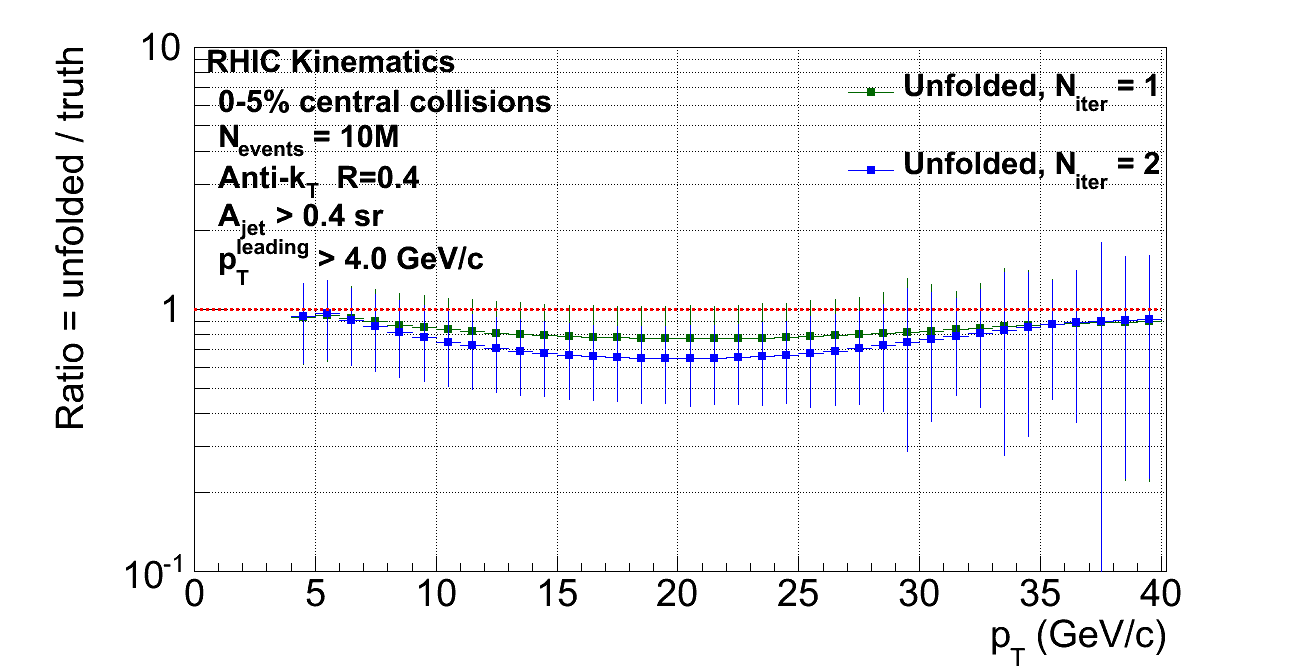}
  \includegraphics[width=0.41\textwidth]{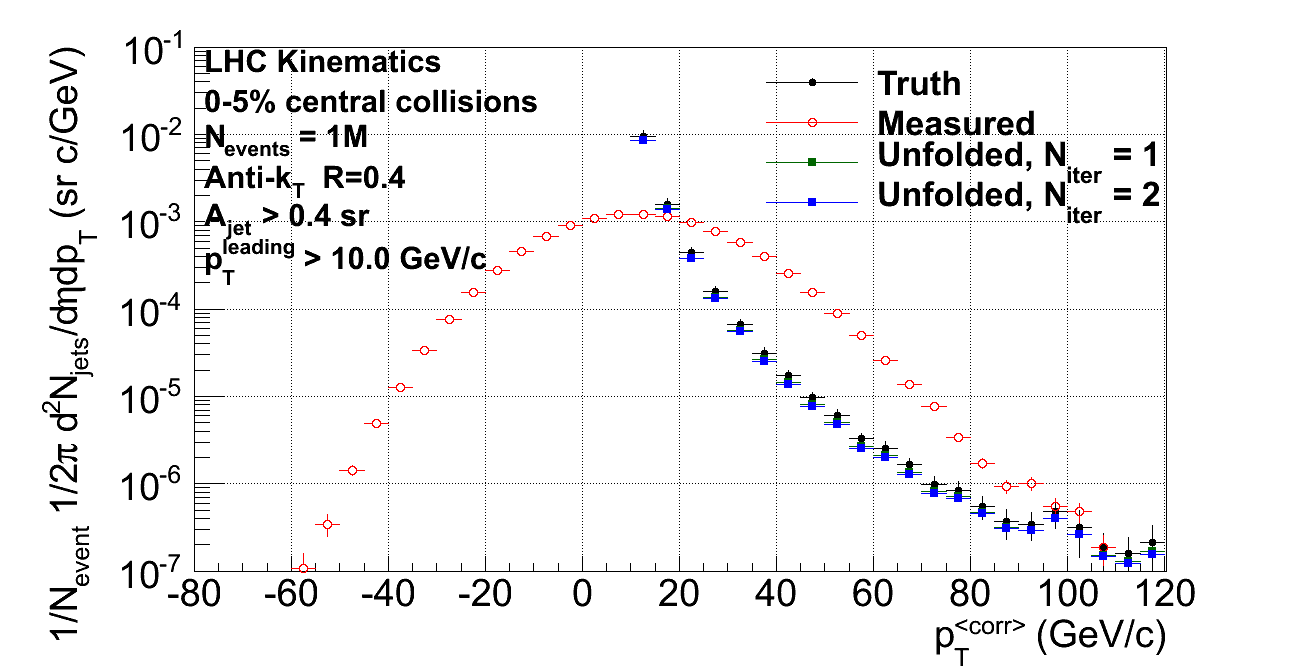}
  \includegraphics[width=0.41\textwidth]{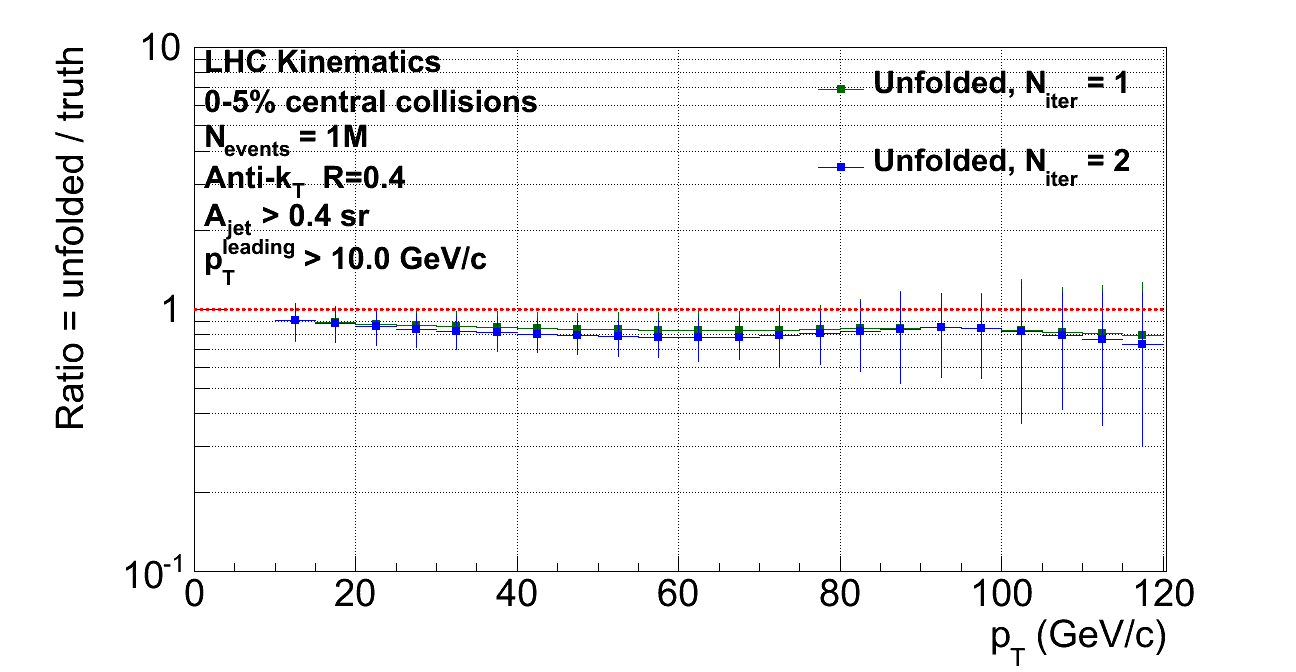}
  \caption{Measured, truth and unfolded distributions for inclusive, jets with 
    leading particle above a given threshold. RHIC (upper plots) and LHC 
    (bottom plots).}
  \label{fig:unf_inc_cut}
\end{figure}

We first consider measurement of the fully inclusive jet
spectrum. Figure \ref{fig:unf_inc_nocut} shows the Truth, measured,
and unfolded jet distributions for RHIC (upper panels) and LHC (bottom
panels). The distributions resulting from the first two Bayesean
unfolding iterations disagree strongly with the Truth distribution,
and successive iterations do not reduce the discrepancy. This effect
is due to the presence of an overwhelming population of combinatorial
background jets which do not have an underlying
physical distribution, but which the unfolding algorithm
cannot distinguish from true jets. Figure \ref{fig:unf_inc_cut} shows the same calculation, but with
the distribution to be unfolded restricted to jets containing a
leading particle $p_{T}^{\rm leading}>4.0$ GeV/c at RHIC and
$p_{T}^{\rm leading}>10.0$ GeV/c at LHC. This cut strongly suppresses the
combinatorial jet background distribution, and the resulting Unfolded
distribution corresponds to the Truth distribution within 10\% for all
$p_{\rm T}$.

The calculations presented here employ simplified SP fragmentation,
though calculations employing PYTHIA fragmentation show similar behavior as
a function of the $p_{T}^{\rm leading}$ threshold. The behavior in data
will depend on the detailed interplay of (quenched) jet fragmentation and
the background hadron distribution, including effects of flow, which
are not included in the Toy Model.

It is apparent that accurate unfolding of background fluctuations from
the inclusive spectrum requires suppression of the combinatorial
background, via imposition of a fragmentation bias (see also
\cite{Hanks:2012wv}). However, this approach also biases the resulting true jet
population, which may be significant for quenching
measurements. The lowest threshold value for $p_{T}^{\rm leading}$ that
enables stable unfolding of the background fluctuations is clearly
preferred, but quantitative assessment of the remaining biases in such
an approach are beyond the scope of this model study.

Given effective suppression of the combinatorial jet population, the
precision of the inclusive measurement is then limited by the precision with
which the unfolding response matrix is known. The response matrix
includes contributions both from detector effects and from the fluctuating
event background, with the latter generally dominant in heavy ion
measurements. Data-driven techniques
\cite{bib:pmj_hp10,bib:gbarros_PANIC11,Abelev:2012ej} are able to
measure the fluctuation distribution $\delta{p_{\rm T}}$ over several
decades in magnitude.

\section{Hadron+Jet Coincidence Measurement}
\label{sec:hjet}

We next turn to a coincidence measurement, the semi-inclusive rate of
jets recoiling against a high $p_{\rm T}$ hadron trigger (``h+jet''). The
acceptance for the recoil jet is

\begin{equation}
  |\varphi_{\rm hadron} - \varphi_{\rm jet} - \pi| < \pi/4.
  \label{eq:deltaphi}
\end{equation}

\noindent
The basic observable is the $p_{\rm T}$ distribution of the number of observed recoil jets, normalized
by the number of triggers. We prefer a hadron to a jet trigger, since hadrons can be accurately
measured in heavy ion collisions without having to account for complex
background effects, and their (suppressed) production is
well-understood in jet quenching models. In addition, model studies
suggest that a high $p_{\rm T}$ hadron trigger imposes a significant
``surface'' bias on the measured population, corresponding to maximum
the path length in matter for recoiling jets, whereas a jet trigger
does not \cite{RenkBiasHadron,RenkBiasJet}.

\begin{figure}[!h]
  \centering
  \includegraphics[width=0.41\textwidth]{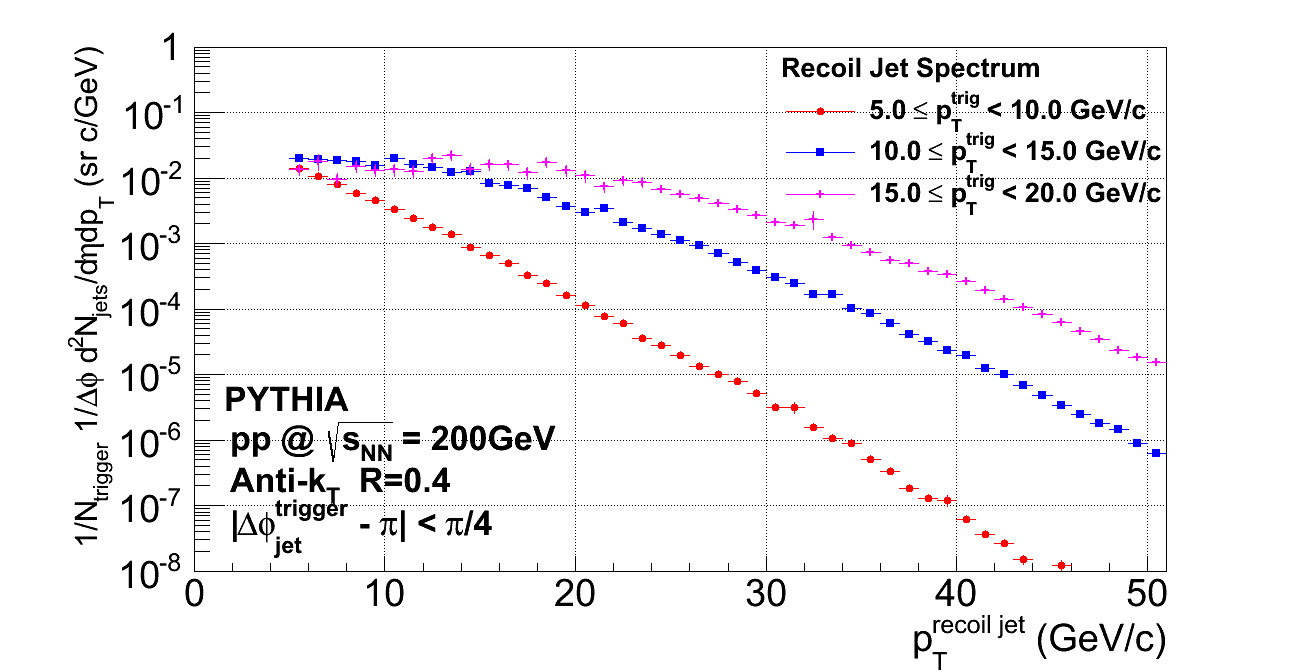}
  \includegraphics[width=0.41\textwidth]{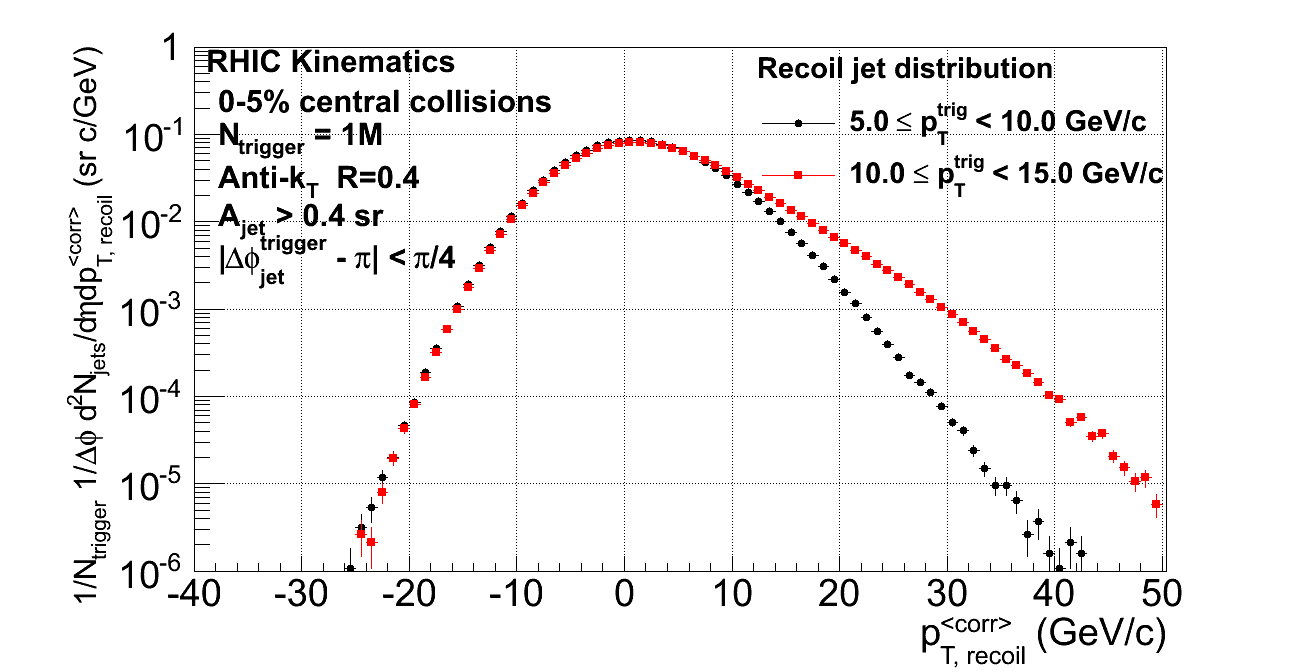}
  \includegraphics[width=0.41\textwidth]{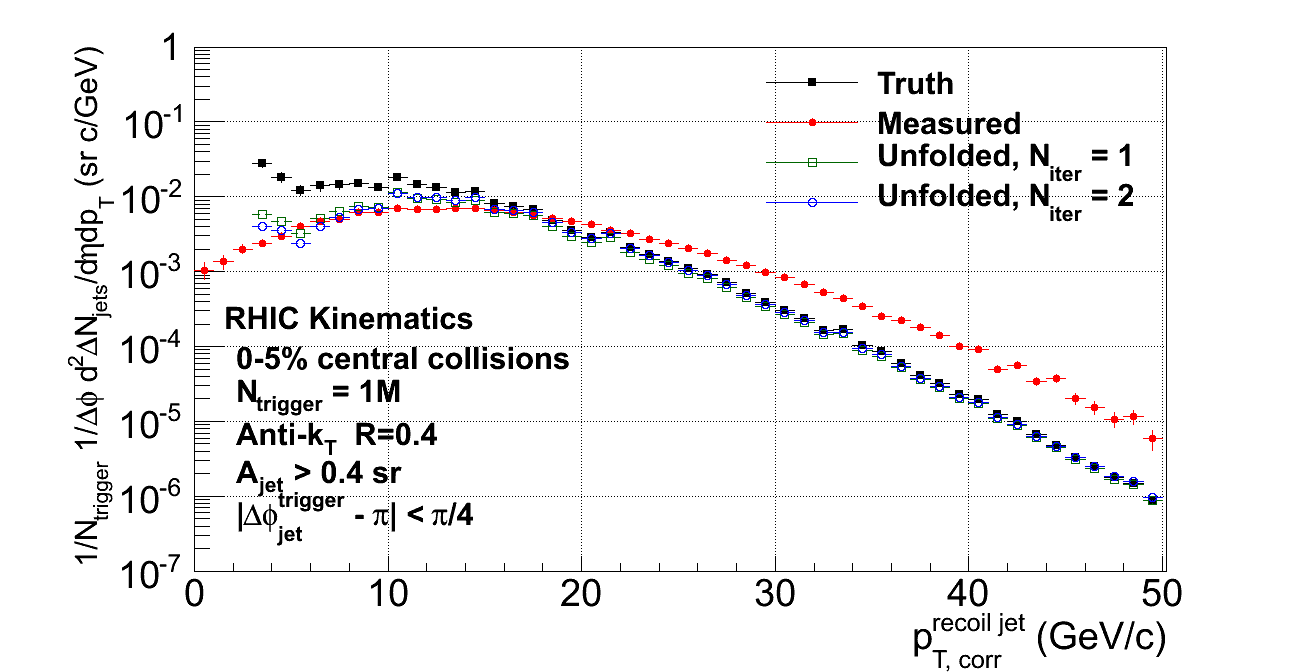}
  \includegraphics[width=0.41\textwidth]{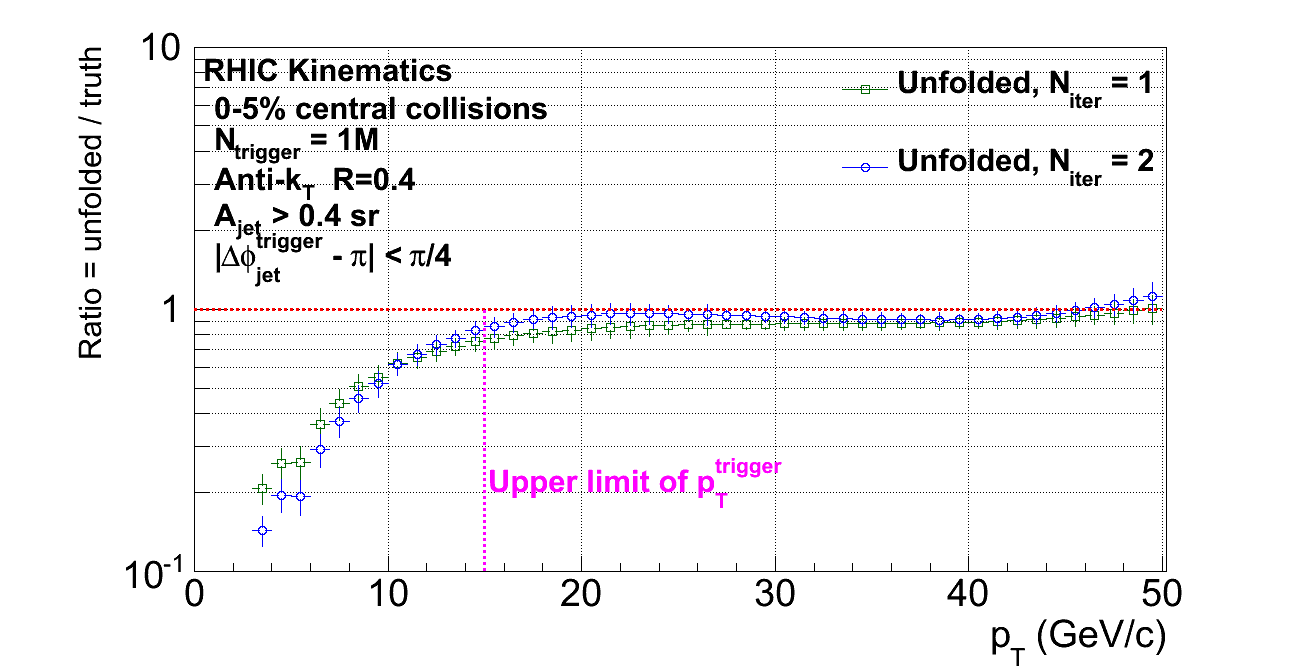}
  \caption{h+jet measurements at RHIC ($\sqrt{s}=200$ GeV) :
    semi-inclusive jet distributions. Error bars are from event
    generation and correspond to very large integrated
    luminosity. Upper left: PYTHIA calculation for p+p collisions with
    various intervals of trigger hadron $p_{\rm T}$. Remaining panels are
    Toy Model calculations for central Au+Au collisions. Upper right:
    $p_{T}^{\left<\rm corr\right>}$ distribution for two intervals of
    trigger hadron $p_{\rm T}$; lower left: difference distribution and
    results of unfolding; lower right: ratio Unfolded/Truth.}
  \label{fig:unf_hjet_RHIC}
\end{figure}

Fig. \ref{fig:unf_hjet_RHIC}, upper left panel, shows a PYTHIA
calculation of the semi-inclusive h+jet distribution for p+p
collisions at RHIC $\sqrt{s}=200$ GeV, for three different intervals
of hadron trigger $p_{\rm T}$. Note that the recoil jet distribution depends
strongly on hadron trigger $p_{\rm T}$, and the probability to
observe a recoil jet in the acceptance with $p_{\rm T}$ above the hadron
trigger threshold is at most a few percent.
 Fig. \ref{fig:unf_hjet_RHIC}, upper right, shows the equivalent recoil
jet distribution for central Au+Au collisions, for two intervals of
trigger hadron $p_{\rm T}$. Unfolding of background fluctuations for
these individual distributions fails in a way similar to unfolding of
fully inclusive distributions (Fig. \ref{fig:unf_inc_nocut}), and for
the same reason: the recoil region is dominated by combinatorial
background jets, even for a high $p_{\rm T}$ trigger, due to the low
probability for true coindiences.  In order to overcome this
background effect without imposing a fragmentation bias on the recoil
jet population, we note that the combinatorial jet distribution is, by
definition, {\it independent} of trigger $p_{\rm T}$. This invariance
is seen in the upper right panel, which compares the recoil
$p_{T}^{\left<\rm corr\right>}$ distribution for two different trigger
$p_{\rm T}$ ranges. The region $p_{T}^{\left<\rm corr\right>}<0$,
where the combinatorial background component may be expected to
dominate, is very similar for the two trigger intervals, whereas the
region for large positive $p_{T}^{\left<\rm corr\right>}>0$ exhibits
strong correlation with trigger $p_{\rm T}$. This invariance may not
be precise in real data analysis, due to biases in reaction plane
orientation and centrality imposed by the requirement of a high
$p_{\rm T}$ trigger hadron. Such biases are not present in Toy Model
events. Their effects in real data analysis can however be minimized,
as described below.

We exploit this invariance by defining a new observable: the {\it
  difference} between the two distributions in
Fig. \ref{fig:unf_hjet_RHIC}, upper right panel, which represents the
{\it evolution} of the recoil distribution with trigger $p_{\rm T}$. The
hadron trigger $p_{\rm T}$ for both ranges should be chosen to be high
enough that the probability per event for such a trigger is low, so
that a hard recoil jet most likely originates from the same hard
interaction as the trigger. In addition, for sufficiently high $p_{\rm T}$
hadron triggers, the reaction plane and centrality biases of the
trigger have weak, if any, dependence on trigger $p_{\rm T}$, and such
biases will affect the combinatorial background similarly for the two
trigger $p_{\rm T}$ ranges. After accounting for the strict conservation of
jet density \cite{bib:gbarros_PANIC11}, the combinatorial background
jet component can be suppressed in the difference distribution in a
purely data-driven way, at the per-mil level of precision or better. Fig. \ref{fig:unf_hjet_RHIC}, lower left, shows the difference
distribution (``Measured'', red points), which represents the hard
recoil jet distribution, but with energy still smeared by background
fluctuations. The lower left panel also shows the result of unfolding
of these fluctuations compared to the Truth distribution from PYTHIA
p+p events, with the ratio Unfolded/Truth shown in lower right
panel. The Truth distribution is recovered within $\sim10\%$ precision
over a broad kinematic range above the trigger threshold, with minimal
cut on the jet constituents ($p_{\rm T}>200$ MeV) and without the
imposition of any fragmentation bias on the measured jet
population. Similar results are obtained for h+jet measurements at the
LHC, with appropriate choice of trigger $p_{\rm T}$.





\bibliographystyle{elsarticle-num}
\bibliography{bibliography}







\end{document}